\documentclass[twocolumn,showpacs,preprintnumbers,amsmath,amssymb]{revtex4}
\begin{document}
\title{Asymptotically Free $\phi^4$ in 3+1}

\author{G. B. Pivovarov}
\email{gbpivo@ms2.inr.ac.ru}
\affiliation{Institute for Nuclear Research,\\
Moscow, 117312 Russia}

\date{October 30, 2005}
 
\begin{abstract}
We demonstrate that regularization with higher derivatives in 3+1 Minkowski space leads to an asymptotically free theory of interacting scalar field.
\end{abstract}

\pacs{11.10.Hi}

\maketitle

As known \cite{Stueckelberg:1953dz}, $\phi^4$ in 3+1 space-time has a Landau pole. Nonperturbatively,
it is related to the statement that there exists only trivial scalar field
\cite{Frohlich:1982tw}. In phenomenology, these assertions lead to the triviality bound on the Higgs
mass \cite{Dashen:1983ts}.

There are searches for nonperturbative constructions that would define nontrivial interacting scalar
fields \cite{Klauder:2000ud}.

Here we demonstrate that there is a possibility to define an interacting scalar field in a way that will imply
(at least perturbatively) asymptotic freedom, and, at the same time, the tree level interaction will be $\phi^4$.

We start with the Lagrangian density for $\phi^4$ amended with a higher derivative regularization term:
\begin{equation}
\label{lagr}
{\cal L} = \frac{1}{2}\partial_\mu\phi\partial^\mu\phi - \frac{m^2}{2}\phi^2-
\frac{g}{4!}\phi^4 -\frac{1}{4 M^2}(\square\phi)^2.
\end{equation}
The square in the past term denotes the d'Alambertian, $\square\phi\equiv\partial_\mu\partial^\mu\phi$.
It is introduced as an ultraviolet regularization. In Euclidean space, this term 
would make (almost) all diagrams finite.

It is a matter of principle to us that we consider the theory with the above Lagrangian in Minkowski space.
With higher derivatives, the possibility to go over to Euclidean space is not granted, and we should consider
the possibility of appearance of extra ultraviolet divergences coming from singularities near the light cone.

As known \cite{Hawking:2001yt}, theories with higher derivatives are plagued with negative norm states, and other
paradoxes. These are not relevant for our consideration, as we consider $M$ to be a sort of ultraviolet cutoff, and 
assume that all the paradoxes disappear in the limit $M\rightarrow\infty$. Thus, the term in (\ref{lagr}) suppressed 
by $M$ introduces a regularization similar to the Pauli-Villars regularization. A crucial difference (see below) is that we introduce it on the level of Lagrangian, so it is defined nonperturbatively.

As we are not certain that the above regularization will make Feynman diagrams finite in Minkovski space, we
will cutoff momentum integrals at the scale $\Lambda$. The idea is that both $M$ and $\Lambda$ scale to infinity,
but $M$ stays much less than $\Lambda$. In particular, we will consider the regime 
\begin{equation}
\label{reg}
M^2=\frac{\Lambda^2}{C\ln(\Lambda/m)}, 
\end{equation}
were $C$ is a positive number.

Now we compute the one loop beta function in the above setting. As we will see, the sign of the
one loop beta function coefficient depends on the value of $C$: If $C>1$, we have asymptotic freedom instead
of the Landau pole.

As known \cite{Weinberg:1996kr}, the standard one loop beta function is 
\begin{equation}
\label{beta}
\beta(g)=\frac{3}{16 \pi^2}\frac{\partial I(\ln\Lambda)}{\partial \ln\Lambda}g^2,
\end{equation}
where $I$ is the one loop Feynman integral
\begin{equation}
\label{int}
I(\ln\Lambda^2)=\frac{1}{2i\pi^2}\int_\Lambda\frac{d^4k}{(k^2-m^2+i\epsilon)^2}.
\end{equation}
Here the integral is defined via rotating to Euclid, and cutting the Euclidean integral by requiring the
Euclidean momentum to satisfy the inequality $k_E^2<\Lambda^2$. Normalization of the integral is tuned to have the derivative of the integral in the above formula equal to unit, so $\beta(g)=3g^2/(16\pi^2)$. [To simplify 
computations, we set the external momentum to zero in the above integral. We checked that the outcome of our consideration can be reproduced for a finite Euclidean external momentum.] 

Now, with the modified Lagrangian (\ref{lagr}), the standard derivation of the beta function leads
to modification of the one loop integral $I$ involved in (\ref{beta}):
\begin{equation}
\label{modint}
I(\ln\Lambda)=\frac{1}{2i\pi^2}\int_\Lambda\frac{d^4 k}{(k^2-m^2+i\epsilon-(k^2)^2/(2 M^2))^2}.
\end{equation}

To compute this integral, we first represent it is a derivative in $m^2$:
\begin{equation}
\label{der}
I=\frac{1}{2i\pi^2}\frac{\partial}{\partial m^2}\int_\Lambda\frac{d^4 k}{k^2-m^2+i\epsilon-(k^2)^2/(2 M^2)}.
\end{equation}
Now we expand the integrand as follows:
\begin{eqnarray}
\label{exp}
\frac{1}{k^2-m^2+i\epsilon-(k^2)^2/(2 M^2)}&=&\frac{1}{\sqrt{1-2m^2/M^2}}\times\nonumber\\
\big[\frac{1}{k^2-\tilde{m}^2+i\epsilon}&-&\frac{1}{k^2-\tilde{M}^2-i\epsilon}\big],
\end{eqnarray}
where
\begin{eqnarray}
\label{tilde_m}
\tilde m^2&\equiv&M^2-\sqrt{M^4-2M^2m^2},\\
\label{tilde_M}
\tilde M^2&\equiv&M^2+\sqrt{M^4-2M^2m^2}.
\end{eqnarray}
Notice that $\tilde m^2\rightarrow m^2$, and $\tilde M^2\rightarrow 2M^2$ when $M^2\rightarrow\infty$.

Now, for both terms in the above expression for the integrand, we can go over to Euclid:
\begin{equation}
\label{inteuclid}
I=-\frac{1}{2\pi^2}\frac{\partial}{\partial m^2}\frac{1}{R}\int_\Lambda\Big[\frac{d^4k}{k_E^2+\tilde m^2}+
\frac{d^4k}{k_E^2+\tilde M^2}\Big],
\end{equation}
where we introduced the notation $R\equiv\sqrt{1-2m^2/M^2}$. Notice that the relative sign of the two terms has changed in Euclid. This is due to the 'wrong' sign of the $i\epsilon$ term in the second term of (\ref{exp}). The integrals are cut by the requirement $k^2_E<\Lambda^2$.

The derivative in $m^2$ in (\ref{inteuclid}) may act on the preintegral factor $1/R$, or on the tilded masses in
the integrand. According to this, we resolve the integral into two terms:
\begin{equation}
\label{two}
I=I_R+I_S,
\end{equation}
where $I_R$ is the term originating from the action of the derivative in $m^2$ on the prefactor $1/R$. The subscript
on the second term means Standard, because this term leads to the standard beta function. Explicitly,
\begin{equation}
\label{i_r}
I_R=-\frac{1}{2\pi^2M^2R^3}\big[\int_\Lambda\frac{d^4k}{k^2_E+\tilde m^2}+
\int_\Lambda\frac{d^4k}{k^2_E+\tilde M^2}\big],
\end{equation}
\begin{equation}
\label{i_e}
I_S=\frac{1}{2\pi^2R}\big[\int_\Lambda\frac{d^4k}{(k^2_E+\tilde m^2)^2}-
\int_\Lambda\frac{d^4k}{(k^2_E+\tilde M^2)^2}\big].
\end{equation}

$I_S$ is finite in the limit $\Lambda \rightarrow \infty$:
\begin{equation}
I_S=\frac{1}{R}\ln\frac{\tilde M^2}{\tilde m^2}.
\end{equation}

Taking into account the regime (\ref{reg}), the expressions for tilded masses, and the definition for 
$R$, we observe that
\begin{equation}
\frac{\partial I_S}{\partial \ln \Lambda} = 1
\end{equation}
in the limit $\Lambda\rightarrow\infty$. Thus, $I_S$ yields the standard beta function. Let us compute
the contribution of $I_R$.

Both integrals in (\ref{i_r}) are quadratically divergent. Keeping only the leading quadratically divergent terms we obtain
\begin{equation}
\label{i_r_expl}
I_R=-\frac{\Lambda^2}{M^2}.
\end{equation}
In the regime (\ref{reg}), we have
\begin{equation}
\frac{\partial I_R}{\partial \ln \Lambda} = -C.
\end{equation}

Thus, we have for the one loop beta function
\begin{equation}
\label{beta_new}
\beta(g)=\frac{3}{16 \pi^2}\big[1-C\big]g^2.
\end{equation}
Recall that $C$ is a positive number in the definition of the regime (\ref{reg}).

As promised, if $C>1$, beta function is negative, and we have asymptotic freedom. At $C=1$, the one loop beta function
vanishes.

We summarize that higher derivatives in Minkowski space change ultraviolet behavior.
Let us stress that if we first go over to Euclidean formulation, and next introduce higher derivatives, the standard
beta function would be reproduced. The way the coupling runs is observable. Thus, the above theory may be either supported or falsified with observations. 

We close this letter with a list of open questions: 
\begin{itemize}
\item 
Is the above theory renormalizable in all orders of perturbation
theory?
\item
What is the one-loop effective action of the above theory, does it imply spontaneous symmetry breaking?
\item
How to give a nonperturbative formulation of this theory?
\end{itemize}

This work was supported in part by RFBR grant no. 03-02-17047.

\end{document}